\UseRawInputEncoding
\documentclass[UTF8]{revtex4}
\usepackage{enumitem}
\usepackage{amsmath,amssymb}
\usepackage{graphicx} 
\usepackage{epstopdf}
\usepackage{hyperref}
\usepackage{geometry}
\usepackage{subfigure}
\usepackage[section]{placeins}
\hypersetup{
    colorlinks=true,
    citecolor=blue,
    linkcolor=red,
    filecolor=magenta,
    urlcolor=cyan,}

\newcommand{\be}{\begin{equation}}
\newcommand{\ee}{\end{equation}}

\begin{document}

\title{Thermodynamics of the Bardeen Black Hole with quintessence matter on the EGUP framework}
\author{Shurui Wu\footnote{gs.srwu20@gzu.edu.cn}$^{1}$, Bing-Qian Wang\footnote{wangbingqian13@yeah.net}$^{2}$, Hao Chen\footnote{gs.ch19@gzu.edu.cn}$^{1}$, and Zheng-Wen Long\footnote{zwlong@gzu.edu.cn (Corresponding author)}$^{1}$}
\affiliation{$^1$ College of Physics, Guizhou University, Guiyang, 550025, China.\\ $^{2}$ College Pharmacy, Guizhou University of Traditional Chinese Medicine, Guiyang 550025, China.
}

\date{\today}

\begin{abstract}
In this paper, we have used the extended generalized uncertainty principle to investigate the thermodynamics of the Bardeen Black Hole with quintessence matter on the EGUP framework. We started with a brief perusal of the Extended generalized uncertainty principle. Subsequently, the EGUP-modified Hawking temperature, heat capacity, and entropy functions of the Bardeen Black Hole are obtained, which show that the modified uncertainty principles and normalization factor restrict the lower bound value of horizon radius to affect the Hawking temperature. Moreover, we compare and discuss the findings within the generalized uncertainty principle (GUP) and extended uncertainty principle (EUP). In addition, we examine the equation of state associated with the pressure and the volume.

\end{abstract}
\maketitle
~~~~~~~~~\textsl{Keywords:}  Dark energy, Bardeen Black Hole, Thermodynamics, Extended generalized uncertainty principle.
\section{Introduction}
Black holes (BHs) are a family of astronomical compact objects, which has a strong gravitational pull that nothing can escape from it not even light. There exist lots of enigmatical physical phenomena related to these compact objects, the most well known one is the presence of the central singularity. General relativity has described the singularity-free BH solutions, which are asymptotically flat and  static spherically symmetric spacetimes labed regular BHs. Bardeen \protect \cite{ret1} firstly presented the singularity-free solution of spherically symmetric BH, and on this basis, Hayward \protect \cite{ret2} developed a nonrotating regular BH, subsequently, Bambi and Modesto \protect \cite{ret3} introduced the family of rotating regular BH solutions after the combination of Bardeen as well as Hayward BHs. Recent astronomical observations indicate that our universe is accelerating expansion \protect \cite{ret4}, and it contains a bulk amount of non-radiating matter distribution named as dark matter, which is composed of unfamiliar subatomic particles. Over the last decade, the study of BHs in the presence of quintessence type dark energy has become an fascinating topic. For example, Kiselev \protect \cite{ret5} firstly proposed symmetrically exact solutions of the Einstein¡¯s equation for black holes surrounded by quintessence matter, Ref. \protect \cite{ret6} studied spherically symmetric BH solution with perfect fluid dark matter by using Newman-Janis algorithm on Kerr-like BH, Ref.\protect \cite{ret7} discussed the Kerr-Newman-anti-de Sitter spacetime in the presence of perfect fluid dark matter, Ref.\protect \cite{ret8} examined the phase transition and thermodynamics for the Reissner Nordstr{\"o}m-anti-de Sitter BH. On the other hand, the thermodynamical properties of BHs are shown to be affected by quintessence matter. For example, S. Chen et al \protect \cite{ret9} examined the Hawking radiation of a static d-dimensional spherically symmetric black hole with quintessence matter, Yuchen Huang et al studied phase structures and transitions of RN black holes surrounded by quintessence dark energy under the condition of AdS space and a Dirichlet wall.

Investigation of the gravitational interaction as an intrinsic quantum nature phenomenon is probably the most challenging and interesting problem of physics. Various approaches are used to study quantum gravity such as string theory, loop quantum gravity and quantum geometry\protect \cite{ret10,ret11,ret12,ret13,ret14}, all of which indicate the existence of a minimal measurable length of the order of the Planck length, therefore the standard Heisenberg uncertainty principle (HUP) has to be modified as Generalized Uncertainty Principle (GUP), and the quantization of gravity motivated by GUP is a an adequate tool for describing black holes and their thermodynamic properties. For example, Refs.\protect \cite{ret15,ret16} studied the thermodynamics of BHs under GUP and obtained the GUP-corrected temperature, entropy and heat capacity, Ref.\protect \cite{ret17} investigated the deformation of the second and third quantized theories by deforming the canonical commutation relations based on GUP, one type of GUP that it is compatible with specific quantum gravity scenarios with a fundamental minimal length and Lorentz violation is derived in Ref. \protect \cite{ret18}, and in Refs. \protect \cite{ret19,ret20}, the effect of a new version of GUP has been investigated on the inflationary cosmology and the dynamics of the Universe respectively.

Motivated by the investigation of minimal measurable length, Kowalski-Glikman et al reported a correction term that is proportional to the radius of de Sitter space-time to characterize a minimum momentum value, this correction is labed the extended uncertainty principle (EUP) \protect \cite{ret21}. Recently, based on the EUP, the authors studied the effects of quantum fluctuations spewed by a black hole and the thermodynamic properties of a black hole \protect \cite{ret22,ret23}. The GUP and EUP are supposed to play an important role in the early and later stages of the universe \protect \cite{ret24}, thus the investigation of the combination of these two extended forms to generate a more general form is worth to be explored, which is referred to as the extended generalized uncertainty principle (EGUP). EGUP can be derived from GUP and EUP that motivated by the quantum-gravity idea of the smallest possible position and by physics in anti-de Sitter space respectively, and it satisfies a large scale spacetime and a wide applicability range. In recent years, the EGUP is employed to discuss the Schwarzschild (anti-)de Sitter black hole thermodynamics \protect \cite{ret25}, thermodynamics of the Friedmann-Robertson-Walker universe \protect \cite{ret26}, the Unruh effects of a black hole \protect \cite{ret27} and thermodynamics of the Reissner-Nordstr\"om black hole with quintessence matter \protect \cite{ret28}.

Motivated by the above works, in this paper, we are interested to study the thermodynamics of the Bardeen Black Hole with quintessence matter on the EGUP framework. The outline of the present work is as follows. In Sect. \protect \ref{sec2}, we present a brief review of the new extended uncertainty principle. The thermodynamics of the Bardeen Black Hole with QM in the context of the EGUP in Sect. \ref{sec3}. Finally, a conclusion is presented in Sect. \protect \ref{sec4}.

\section{A brief review of the new extended uncertainty principle} \label{sec2}
In GUP the Heisenberg¡¯s uncertainty principle receives a correction term from gravitational effect
\begin{equation} \Delta x_{i} \Delta p_{j} \geq \frac{\hbar }{2} \delta_{i j}\left[1+\frac{\beta l_{p}^{2}}{\hbar^{2}}\left(\Delta p_i \right)^{2} \right], i,j=1,2,3. \label{eq1} \end{equation}
here $x_{i}$, $p_{j}$, $\beta$ and $l_p$ denote the spatial coordinate,  momentum operators, the dimensionless GUP parameter of order unity and the Planck length respectively, and it implies an observable minimal length as $\Delta x_{i}\geq \sqrt{\beta} l_{p}$. Moreover, we can obtain an uncertainty range of momentum in terms of Eq.\eqref{eq1} as
\begin{equation}
\frac{\hbar \Delta x_{j}}{\beta l_{p}^{2}}\left[1-\sqrt{1-\frac{ \beta l_{p}^{2}}{\left(\Delta x\right)^{2}}}\right] \leq \Delta p_{j} \leq \frac{\hbar \Delta x_{j}}{\beta l_{p}^{2}}\left[1+\sqrt{1-\frac{ \beta l_{p}^{2}}{\left(\Delta x\right)^{2}}}\right], \label{eq2}   \end{equation}
as we stated before, the GUP is supposed to play an important role in the early universe, thus we can comprehend Eq.\protect \eqref{eq2} as: the correction term contributes to the uncertainty significantly if and only if the position uncertainty is of the order of the Planck length. Recently, Z.L. Zhao et al proposed a new high-order GUP in the form of \protect \cite{ret29}
\begin{equation}
\Delta x_{i} \Delta p_{j} \geq \frac{\hbar \delta_{i j}}{2} \frac{1}{1-\beta\left[(\Delta p)^{2}-\langle p\rangle^{2}\right]}, \label{eq3}  \end{equation}
in which a minimal uncertainty of position is reported as $\frac{3 \sqrt{3}}{4} \hbar \sqrt{\beta}$. While unlike the GUP, the EUP formalism is believed to play an important role in the latter stages of the universe, it's formalism is given by
\begin{equation} \Delta X_{i} \Delta P_{j} \geq \frac{\hbar \delta_{i j}}{2}\left[1+\alpha_{0}^{2}\left(\Delta X_{i}\right)^{2}\right], \label{eq4}  \end{equation}
here the deformation parameter satisfies $\alpha_{0}^{2}= 1 / l_{H}^{2}$, and $l_{H}$ is the (anti-)de Sitter space-time radius. In this work, we consider the following EGUP formalism as a linear superposition of the GUP and EUP
\begin{equation}
\Delta x_{i} \Delta p_{j} \geq \frac{\hbar \delta_{i j}}{2}\left[1+\beta l_{p}^{2} \frac{\left(\Delta p _i \right)^{2}}{\hbar^{2}}+\alpha \frac{\left(\Delta x_i \right)^{2}}{l^{2}}\right],
\label{eq5}  \end{equation}
where $\beta$ and $\alpha$ are positive deformation parameters that depend on the expectation values of $x$ and $p$, and $l$ represents the large scale quantity associated with space-time radius. In this case, the linear superposition in Eq.\protect \eqref{eq5} will lead to the minimum position and momentum physical values simultaneously as
\small
\begin{equation}
\frac{(\Delta P) l^{2}}{\hbar \alpha}\left[1-\sqrt{1-\frac{\hbar^{2} \alpha}{l^{2}}\left(\frac{1}{(\Delta P)^{2}}+\frac{\beta l_{p}^{2}}{\hbar^{2}}\right)}\right] \leq \Delta X \leq \frac{(\Delta P) l^{2}}{\hbar \alpha}\left[1+\sqrt{1-\frac{\hbar^{2} \alpha}{l^{2}}\left(\frac{1}{(\Delta P)^{2}}+\frac{\beta l_{p}^{2}}{\hbar^{2}}\right)}\right],  \label{eq6}
\end{equation}\normalsize
\begin{equation}
\frac{\hbar(\Delta X)}{\beta l_{p}^{2}}\left[1-\sqrt{1-\beta l_{p}^{2}\left(\frac{1}{(\Delta X)^{2}}+\frac{\alpha}{l^{2}}\right)}\right] \leq \Delta P \leq \frac{\hbar(\Delta X)}{\beta l_{p}^{2}}\left[1+\sqrt{1-\beta l_{p}^{2}\left(\frac{1}{(\Delta X)^{2}}+\frac{\alpha}{l^{2}}\right)}\right].  \label{eq7}
\end{equation}

\section{Thermodynamics of the Bardeen Black Hole with quintessence matter in the context of the EGUP} \label{sec3}

According to the the solution of Einstein's field equations associated with a static spherically symmetric black hole embedded in quintessence matter, the general form of the metric is given by
\begin{equation}
ds^{2}=-f(r) d t^{2}+\frac{1}{f(r)} d r^{2}+r^{2} d \theta^{2}+r^{2} \sin ^{2} \theta d \phi^{2},  \label{eq8}   \end{equation}
where
\begin{equation}
f(r)=1-\frac{2 M}{r}+\frac{3Mg^{2}}{r^{3}}-\frac{c}{r^{3 \omega_{q}+1}}, \label{eq9}    \end{equation}
here $M$, $g$ and $c$ are the mass of the Bardeen black hole, the magnetic charge of the nonlinear self-gravitating monopole and the positive normalization factor associated with quintessence matter respectively. Accelerating expansion of the universe is one of the most interesting results of observational cosmology, this result can be explained by the concept of dark energy. Dark energy can be regarded as an exotic scalar field with a large negative pressure, and which constitutes about 70 percent of the total energy of the universe. Although the nature of the Dark energy is yet to be understood so far, there are some cosmological models proposed in which the dominant component of the energy density has negative pressure. For example,  the cosmological constant corresponds to the case of dark energy with the barotropic index $\omega_{q}=-1$, the free quintessence matter generates the black hole horizon with the barotropic index is $-1<\omega_{q}<0$, the (anti-)de Sitter radius with the barotropic index is $-1<\omega_{q}<-2/3$ and the quintessence regime of dark energy with the barotropic index is $\omega_{q}=-2/3$. It should be noted that one can obtain the Reissner-Nordstr\"om black hole if take the parameter to satisfy $g=0$ and the Schwarzschild black hole can be obtained by demanding $g=0$ and $c=0$ simultaneously.

Before proceeding any further, it should be noted that in this manuscript, we use natural units. According to the Eq.\protect \eqref{eq9}, the event horizon can be expressed as
\begin{equation}\left.\left(1-\frac{2 M}{r}+\frac{3 M g^{2}}{r^{3}}-\frac{c}{r^{3 w_{q}+1}}\right)\right|_{r=r_{H}}=0  , \label{eq10}    \end{equation}
and in the semi-classical case, the Hawking temperature T is
\begin{equation}  T=\frac{\kappa}{8\pi} \frac{dA}{dS}, \label{eq11}   \end{equation}
where $S$ and $A$ represent the entropy and the black hole area respectively, and the surface gravity $\kappa$ is in the form of
 \begin{equation}
\kappa=-\lim _{r \rightarrow r_{H}} \sqrt{-\frac{g^{11}}{g^{00}}} \frac{\left(\left(g^{00}\right)\right)}{g^{00}}^{'}=\frac{1}{r_{H}}\left(1-6g^{2}\frac{1-\frac{c}{r^{3 \omega_{q}+1}_{H}}}{2r^{2}_{H}-3g^{2}}+\frac{3 \omega_{q}c}{r_{H}^{3 \omega_{q}+1}}\right). \label{eq12}  \end{equation}.
Motivated by a heuristic method \protect \cite{ret30}, we supposed that a particle is absorbed by black hole in the vicinity of its horizon, which will result in a minimal increase in the black hole area that is proportional to the product of the position and momentum uncertainties. And the Ref. \protect \cite{ret31} indicates that a minimal change in the entropy which can not be smaller than $\ln 2$, then we rewrite the Hawking temperature as
\begin{equation}
T  \simeq \frac{\kappa \epsilon}{8 \pi \ln 2} \Delta x \Delta p , \label{eq13}
\end{equation}
where $\epsilon$ denotes the calibration factor that ensures the consistency of the obtained result with the semi-classical result, thus we can derive the following modified Hawking temperature by setting $\Delta x\simeq2r_{h}$
\begin{equation}
T=\frac{\hbar \epsilon r_{H}}{2 \pi \beta l_{p}^{2} \ln 2}\left(1-6g^{2}\frac{1-\frac{c}{r^{3 \omega_{q}+1}_{H}}}{2r^{2}_{H}-3g^{2}}+\frac{3 \alpha \omega_{q}}{r_{h}^{3 \omega_{q}+1}}\right)\left[1-\sqrt{1-\frac{\beta l_{p}^{2}}{4 r_{H}^{2}}\left(1+\frac{4 \alpha r_{H}^{2}}{l^{2}}\right)}\right].  \label{eq14} \end{equation}
Obviously, if we take the normalization factor, the magnetic charge of the nonlinear self-gravitating monopole and the deformation parameters as $\alpha=\beta=c=g=0$, we have $T=\epsilon /({16 \pi r_{h} \ln 2})$, and the well-known Hawking temperature $T=1 / ({4\pi r_{H}})$ will appear if we select the calibration factor as $\epsilon = 4 \ln 2 / \hbar $. Subsequently, we express the EGUP modified Hawking temperature associated with the Bardeen Black Hole surrounded with the quintessence matter as
\begin{equation} T=\frac{2r_{H}}{\pi \beta l_{p}^{2} \ln 2}\left(1-6g^{2}\frac{1-\frac{c}{r^{3 \omega_{q}+1}_{H}}}{2r^{2}_{H}-3g^{2}}+\frac{3 \alpha \omega_{q}}{r_{H}^{3 \omega_{q}+1}}\right)\left[1-\sqrt{1-\frac{\beta l_{p}^{2}}{4 r_{H}^{2}}\left(1+\frac{4 \alpha r_{H}^{2}}{l^{2}}\right)}\right].  \label{eq15} \end{equation}
Next, we present a brief discussion in terms of the deformed Hawking temperature. Firstly, by setting $\alpha=0$, we obtain the GUP modified Hawking temperature
\begin{equation} T_{GUP}=\frac{2r_{H}}{\pi \beta l_{p}^{2} \ln 2}\left(1-6g^{2}\frac{1-\frac{c}{r^{3 \omega_{q}+1}_{H}}}{2r^{2}_{H}-3g^{2}}+\frac{3c \omega_{q}}{r_{H}^{3 \omega_{q}+1}}\right)\left[1-\sqrt{1-\frac{\beta l_{p}^{2}}{4 r_{H}^{2}}}\right],  \label{eq16} \end{equation}
and the EUP modified Hawking temperature arises for $\beta=0$
 \begin{equation} T_{EUP}=\frac{1}{4 \pi r_{H}}\left(1-6g^{2}\frac{1-\frac{c}{r^{3 \omega_{q}+1}_{H}}}{2r^{2}_{H}-3g^{2}}+\frac{3 \alpha \omega_{q}}{r_{H}^{3 \omega_{q}+1}}\right)\left[1-\sqrt{1+\frac{4r^{2}_{H}\alpha}{l^{2}}} \right],  \label{eq17} \end{equation}
 finally, we can obtain the HUP modified Hawking temperature by setting $\alpha=\beta=0$
 \begin{equation} T_{HUP}=\frac{1}{4 \pi r_{H}}\left(1-6g^{2}\frac{1-\frac{c}{r^{3 \omega_{q}+1}_{H}}}{2r^{2}_{H}-3g^{2}}+\frac{3 \alpha \omega_{q}}{r_{H}^{3 \omega_{q}+1}}\right).  \label{eq18} \end{equation}

On the other hand, it should be noted that the Hawking temperature must be positive and real-valued, i.e. we need to take a constraint on Eq.\protect \eqref{eq15}. Specifically, the quintessence matter depends on the normalization factor, the magnetic charge of the nonlinear self-gravitating monopole and the barotropic index, which guarantee the positive values, and the other terms with the EGUP parameters ensure to take real values. Next, we present a discussion of the Hawking temperature in terms of the constraint conditions. For example, if we take the barotropic index as $\omega_{q}=-\frac{1}{3}$, then the following constraint on the radius arises
 \begin{equation}
1-6 g^{2} \frac{(1-c)}{2 r_{H}^{2}-3 g^{2}}-c \geq 0,   \label{eq19}  \end{equation}
with $\left(0 \leq c \leq 1, r_{H} \geq \frac{3}{\sqrt{2}}|g|\right)$. While if we takes $\omega_{q}=-\frac{2}{3}$, then the following third-order equation can be obtained as
 \begin{equation}
 -4 c r_{H}^{3}+2 r_{H}^{2}+12 g^{2} c r_{H}-9 g^{2} \geq 0 ,   \label{eq20}  \end{equation}
Obviously, there are three roots responsible for the Eq.\protect \eqref{eq20}, one real root and two imaginary roots. Here, we just present the meaningful real root
\begin{equation}
\begin{split}
r_{H_{r e}}=\frac{1}{6 c}+\frac{-4-144 c^{2} g^{2}}{24 c\left(-1+189 c^{2} g^{2}+9 \sqrt{3} c g \sqrt{-2+131 c^{2} g^{2}-192 c^{4} g^{4}}\right)^{\frac{1}{3}}} \\
-\frac{\left(-1+189 c^{2} g^{2}+9 \sqrt{3} c g \sqrt{-2+131 c^{2} g^{2}-192 c^{4} g^{4}}\right)^{\frac{1}{3}}}{6 c}. \label{eq21} \end{split} \end{equation}
As another case, if we consider $\omega_{q}=-1$, then the following fourth-order equation can be derived as
\begin{equation} -6 c r_{H}^{4}+\left(2+15 g^{2} c\right) r_{H}^{2}-9 g^{2} \geq 0, \label{eq22} \end{equation}
with the following roots:
\begin{equation}  r_{H 1}=\pm \frac{\sqrt{2+15 g^{2} c-\sqrt{4-156 g^{2} c+225 c^{2} g^{4}}}}{2 \sqrt{3 c}}  , \label{eq23} \end{equation}
\begin{equation}
   r_{H 2}=\pm \frac{\sqrt{2+15 g^{2} c+\sqrt{4-156 g^{2} c+225 c^{2} g^{4}}}}{2 \sqrt{3 c}}. \label{eq24} \end{equation}
On the other hand, we take an algebraic analysis on the second constraint in terms of the deformed Hawking temperature, which ensures the real values, i.e.
\begin{equation}
1-\beta l_{p}^{2}\left(\frac{1}{4 r_{H}^{2}}+\frac{\alpha}{l^{2}}\right) \geq 0, \label{eq25} \end{equation}
this constraint condition will produce a lower bound on the horizon
\begin{equation} r_{H} \geq \frac{l_{p} l}{2} \sqrt{\frac{\beta}{l^{2}-\alpha \beta l_{p}^{2}}}. \label{eq26} \end{equation}
Obviously, the constraint on the horizon radius will reduce to $r_{H} \geq \frac{l_{p}}{2} \sqrt{\beta}$ if we consider the GUP scenario, while this constraint will vanish in terms of the EUP and HUP scenarios, and the horizon radius reduces to $r_H\geq 0$.

\begin{figure}[tbh!]
\subfigure{
\includegraphics[width=7cm]{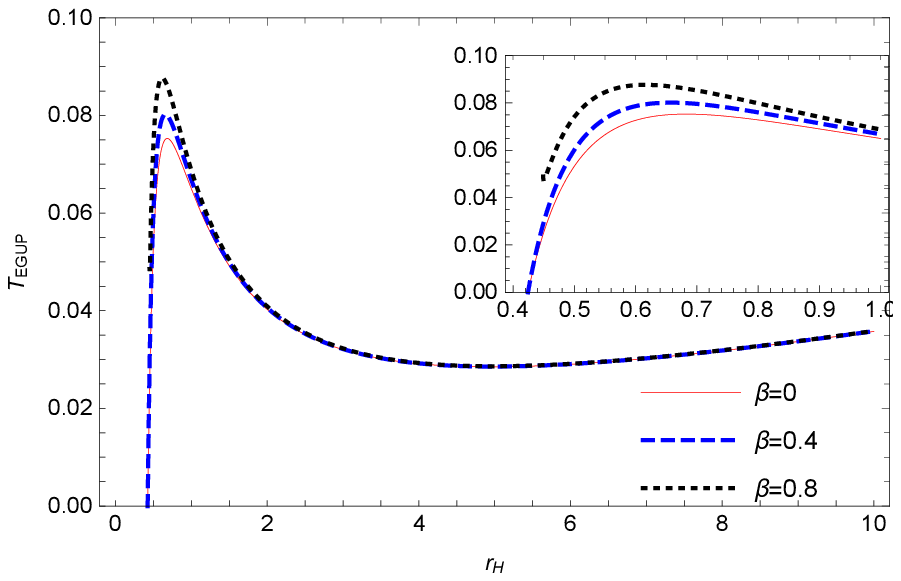}}
\subfigure{
\includegraphics[width=7cm]{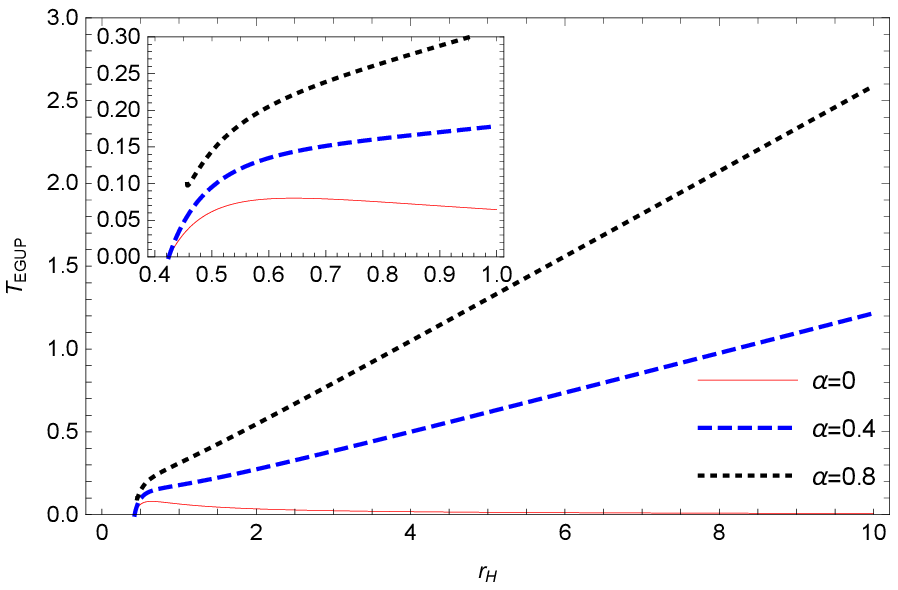}}
\caption{(left panel) Hawking temperature $T_-\frac{1}{3}$ as a function of horizon radius $r_H$ for different GUP parameters, $\beta$, ($\omega_q=-1/3$, $\alpha=0.01$). (right panel) Hawking temperature $T_-\frac{1}{3}$ as a function of horizon radius $r_H$ for different EUP parameters, $\alpha$, ($\omega_q=-1/3$, $\beta=0.5$).}
\label{fig1}
\end{figure}

\begin{figure}
\subfigure{
\includegraphics[width=7cm]{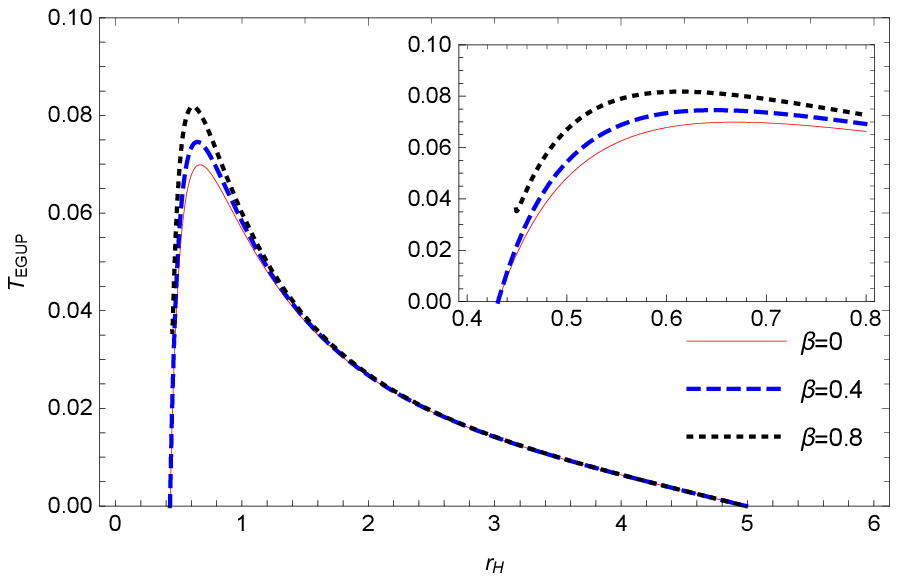}}
\subfigure{
\includegraphics[width=7cm]{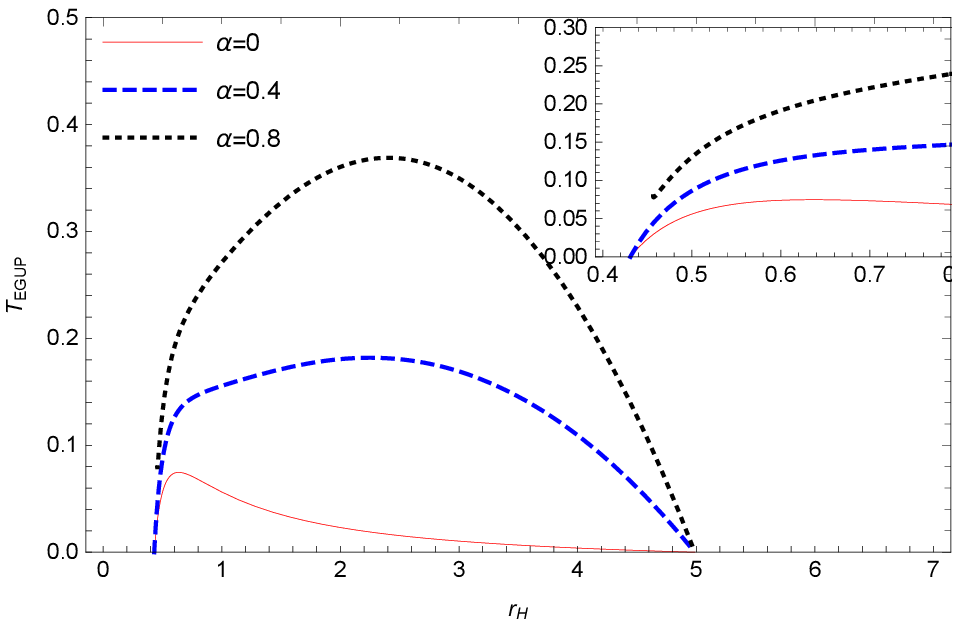}}
\caption{(left panel) Hawking temperature $T_-\frac{2}{3}$ as a function of horizon radius $r_H$ for different GUP parameters, $\beta$, ($\omega_q=-2/3$, $\alpha=0.01$). (right panel) Hawking temperature $T_-\frac{2}{3}$ as a function of horizon radius $r_H$ for different EUP parameters, $\alpha$, ($\omega_q=-2/3$, $\beta=0.5$).}
\label{fig2}
\end{figure}

\begin{figure}
\subfigure{
\includegraphics[width=7cm]{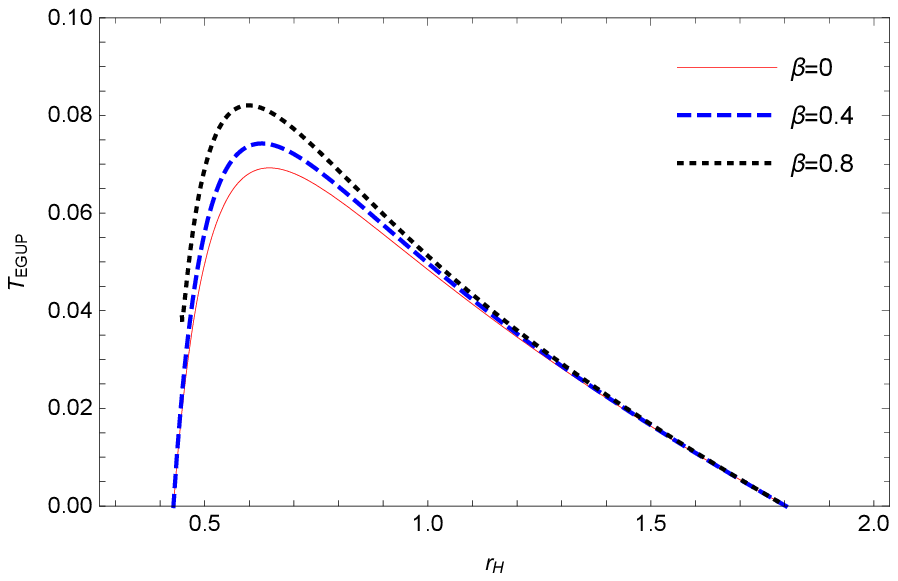}}
\subfigure{
\includegraphics[width=7cm]{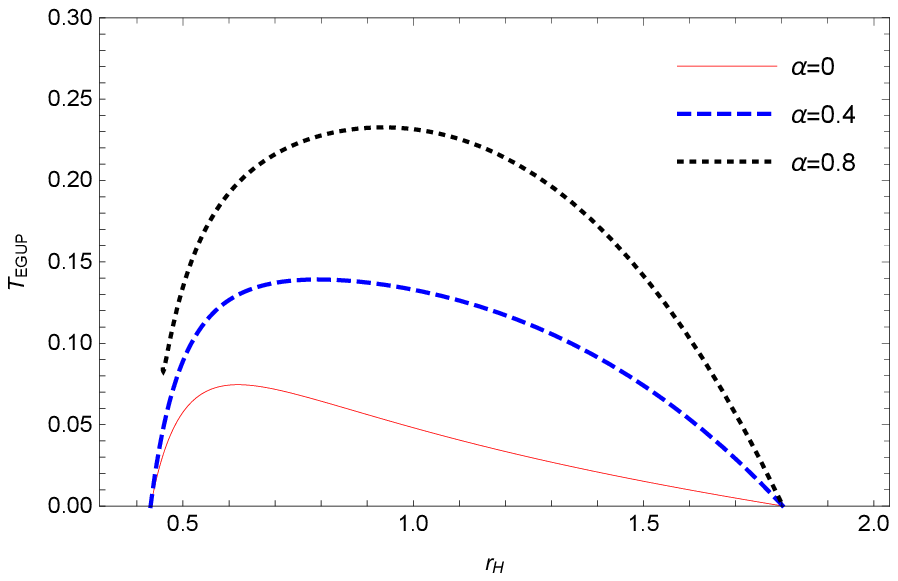}}
\caption{(left panel) Hawking temperature $T_-1$ as a function of horizon radius $r_H$ for different GUP parameters, $\beta$, ($\omega_q=-1$, $\alpha=0.01$). (right panel) Hawking temperature $T_-1$ as a function of horizon radius $r_H$ for different EUP parameters, $\alpha$, ($\omega_q=-1$, $\beta=0.5$).}
\label{fig3}
\end{figure}

In order to facilitate the analysis on the EGUP modified Hawking temperature, we depict it versus horizon for $\omega_q=-1/3$, $\omega_q=-2/3$ and $\omega_q=-1$ 3 in Fig.\ref{fig1}, Fig.\ref{fig2} and Fig.\ref{fig3} respectively. Before proceeding any further, it should be noted that in this manuscript, we set $l=l_p=1$,$c=0.1$ and $g=0.2$. And as stated before, the first constraint condition depends on the values of  the normalization factor, the magnetic charge of the nonlinear self-gravitating monopole and the barotropic index, thus we have the following conditions:

\begin{enumerate}[label=\roman*.]
\item For $\omega_q=-1/3$, Eq. \eqref{eq19} presents $r_H \geq 0.4243$.

\item For $\omega_q=-2/3$, Eq. \eqref{eq20} restricts the horizon to be in the interval, $0.431\leq r_H \leq 4.98$.

\item For $\omega_q=-1$, Eq. \eqref{eq22} gives the interval, $0.430\leq r_H \leq 1.802$.

\end{enumerate}

Next, we present a discussion for the first constraint condition in terms of the $\omega_q=-1/3$ case. For example, we analyse the effect of the GUP parameter by configurating three different GUP parameters and a constant EUP parameter($\alpha=0.01$) in Fig.\protect \ref{fig1}(left panel). Moreover, by considering the second constraint expressed in Eq. \eqref{eq25} in terms of the $\omega_q=-1/3$ case, we have
\begin{enumerate}[label=\roman*.]

\item For $\beta=0$,  $r_H \geq 0$.

\item For $\beta=0.4$,  $r_H \geq 0.317$.

\item For $\beta=0.8$,  $r_H \geq 0.450$.
\end{enumerate}
Since we need to take the intersection range of the two constraint conditions, thus one can summarize the characteristic of the horizon range and Hawking temperature
\begin{enumerate}[label=\roman*.]

\item For $\beta=0.0$, and $\alpha=0.01$,  $r_H \geq 0.4243$ and $T_{EGUP}(0.4243)=0$.

\item For $\beta=0.4$,  and $\alpha=0.01$,  $r_H \geq 0.4243$ and $T_{EGUP}(0.4243)=0$.

\item For $\beta=0.8$, and $\alpha=0.01$,  $r_H \geq 0.450$ and $T_{EGUP}(0.450)=0.0475$.
\end{enumerate}
Obviously, the Fig.\protect \ref{fig1}(left panel) confirms our prediction, which indicates that the GUP scenario is effective in the early stage.

Subsequently, we discuss the effect of the EUP parameter by configurating three different EUP parameters and a constant GUP parameter($\beta=0.5$) in Fig.\ref{fig1}(right panel). And according to the second constraint, we have
\begin{enumerate}[label=\roman*.]

\item For $\alpha=0$,  $r_H \geq 0.354$.

\item For $\alpha=0.4$,  $r_H \geq 0.3953$.

\item For $\alpha=0.8$,  $r_H \geq 0.46$.
\end{enumerate}
As we discussed above, the solutions must be in the intersection range, thus we have
\begin{enumerate}[label=\roman*.]

\item For $\beta=0.5$, and $\alpha=0$,  $r_H \geq 0.4243$ and $T_{EGUP}(0.4243)=0$.

\item For $\beta=0.5$,  and $\alpha=0.4$,  $r_H \geq 0.4243$ and $T_{EGUP}(0.4243)=0$.

\item For $\beta=0.5$, and $\alpha=0.8$,  $r_H \geq 0.46$ and $T_{EGUP}(0.46)=0.01$.
\end{enumerate}
As expected, the plot of Fig.\protect \ref{fig1}(right panel) verifies our findings, which indicates that the EUP scenario is effective in the late stages.

Next, we present a similar analysis for $\omega_q=-2/3$ and $\omega_q=-1$ quintessence matter scenarios respectively. Firstly, for $\omega_q=-2/3$ case, in order to analyze the effect of the GUP parameter, we take a constant EUP parameter $\alpha=0.01$ and three different GUP parameters $\beta=0$, $\beta=0.4$ and $\beta=0.8$ in Table \protect \ref{table1}, and considering the combination of the two constraint conditions, we have
\begin{table}[tbh!]
    \centering
    \begin{tabular}{|c|c|c|c|c|}
    \hline
        $\beta$ & $\alpha$ & horizon range & $T(r_{H_{1}})$ & $T(r_{H_{2}})$ \\
        \hline
         0.0 & 0.01 & $0.431\leq r_H \leq 4.98$ & 0 & 0 \\ \hline 0.4 & 0.01 & $0.431\leq r_H \leq 4.98$ & 0 & 0 \\\hline
         0.8 & 0.01 & $0.45\leq r_H \leq 4.98$ & 0.0354 & 0 \\ \hline
    \end{tabular}
    \caption{The effect of GUP parameter in the quintessence regime of dark energy in terms of $\omega_q=-2/3$.}
    \label{table1}
\end{table}
Similarly, in this case, in order to analyze the effect of the EUP parameter, we take a constant GUP parameter $\beta=0.5$ and three different EUP parameters $\alpha=0$, $\alpha=0.4$ and $\alpha=0.8$ in Table \protect \ref{table2}, and still considering the combination of the two constraint conditions, we obtain
\begin{table}[tbh!]
    \centering
    \begin{tabular}{|c|c|c|c|c|}
    \hline
        $\beta$ & $\alpha$ & horizon range & $T(r_{H_{1}})$ & $T(r_{H_{2}})$ \\
        \hline
         0.5 & 0 & $0.431\leq r_H \leq 4.98$ & 0 & 0 \\
         \hline 0.5 & 0.4 & $0.431\leq r_H \leq 4.98$ & 0 & 0.0011 \\
         \hline
         0.5 & 0.8 & $0.46\leq r_H \leq 4.98$ & 0.081 & 0.0023 \\
         \hline
    \end{tabular}
    \caption{The effect of EUP parameter in the quintessence regime of dark energy $\omega_q=-2/3$.}
    \label{table2}
\end{table}
Finally, for $\omega_q=-1$ case, for the sake of convenience, we just give the corresponding results in Table \ref{table3} and Table \protect \ref{table4}.
\begin{table}[tbh!]
    \centering
    \begin{tabular}{|c|c|c|c|c|}
    \hline
        $\beta$ & $\alpha$ & horizon range & $T(r_{H_{1}})$ & $T(r_{H_{2}})$ \\
        \hline
         0.0 & 0.01 & $0.430\leq r_H \leq 1.802$ & 0 & 0 \\ \hline 0.4 & 0.01 & $0.430\leq r_H \leq 1.802$ & 0 & 0 \\\hline
         0.8 & 0.01 & $0.450\leq r_H \leq 1.802$ & 0.0382 & 0 \\ \hline
    \end{tabular}
    \caption{The effect of GUP parameter in the quintessence regime of dark energy in terms of $\omega_q=-1$.}
    \label{table3}
\end{table}

\begin{table}[tbh!]
    \centering
    \begin{tabular}{|c|c|c|c|c|}
    \hline
        $\beta$ & $\alpha$ & horizon range & $T(r_{H_{1}})$ & $T(r_{H_{2}})$ \\
        \hline
         0.5 & 0 & $0.43\leq r_H \leq 1.802$ & 0 & 0 \\
         \hline 0.5 & 0.4 & $0.43\leq r_H \leq 1.802$ & 0 & 0 \\
         \hline
         0.5 & 0.8 & $0.46\leq r_H \leq 1.802$ & 0.083 & 0 \\
         \hline
    \end{tabular}
    \caption{The effect of EUP parameter in the quintessence regime of dark energy $\omega_q=-1$.}
    \label{table4}
    \end{table}

Next, we express the heat capacity function on the EGUP formalism as

\begin{equation} C_{EGUP}=\frac{\pi r_{H}^{1+3 w_{q}}\left[-9 g^{2} r_{H}^{2}+2 r_{H}^{4}+3 c g^{2}\left(2-3 w_{q}\right) r_{H}^{1-3 w_{q}}+6 c w_{q} r_{H}^{3-3 w_{q}}\right]}{\xi_{1}+\xi_{2}},
\label{eq27} \end{equation}
with
\begin{equation}
\xi_{1}=-\frac{\left(-3 g^{2}+2 r_{H}^{2}\right)\left[3 c g^{2}\left(2-3 w_{q}\right)+6 c w_{q} r_{H}^{2}+r_{H}^{1+3 w_{q}}\left(-9 g^{2}+2 r_{H}^{2}\right)\right]}{2 r_{H}^{2} \sqrt{1-\beta l_{p}^{2}\left(\frac{\alpha}{l^{2}}+\frac{1}{4 r_{H}^{2}}\right)}}
,\label{eq28} \end{equation}

 \begin{equation}
 \begin{split}
\xi_{2}=\frac{1}{\beta l_{p}^{2}}\{2(-1+\sqrt{1-\beta l_{p}^{2}(\frac{\alpha}{l^{2}}+\frac{1}{4 r_{H}^{2}})}[-r_{H}^{1+3 w_{q}}(27 g^{4}+4 r_{H}^{4})+3 c(9 g^{4} w_{q}(-2+3 w_{q}) \\ +4g^{2} r_{H}^{2}(2+3 w_{q}-9 w_{q}^{2})+12 w_{q}^{2} r_{H}^{4})]\}
,\label{eq29} \end{split}
\end{equation}

Since the black hole cannot exchange radiation with its surrounding space when the heat capacity is absence, and the remnant of  black hole is formed. Thus we conclude that the black hole remnant occurs  when the horizon radius demands
$r_{H}=\frac{l_{p} l}{2} \sqrt{\frac{\beta}{l^{2}-\alpha \beta l_{p}^{2}}}$, then we can express the remnant temperature
 \begin{equation}
T_{r e m-E G U P}=\frac{l}{\pi l_{p} \sqrt{\beta\left(l^{2}-\alpha \beta l_{p}^{2}\right)}}\left[1-12 g^{2} \frac{\left(1-c\left(\frac{l_{p} l}{2} \sqrt{\frac{\beta}{l^{2}-\alpha \beta l_{p}^{2}}}\right)^{-\left(3 w_{q}+1\right)}\right)\left(l^{2}-\alpha \beta l_{p}^{2}\right)}{l^{2} l_{p}^{2} \beta-6 g^{2}\left(l^{2}-\alpha \beta l_{p}^{2}\right)}+\kappa\right], \label{eq30}
\end{equation}
with $\kappa=3 w_{q} c\left(\frac{l_{p} l}{2} \sqrt{\frac{\beta}{l^{2}-\alpha \beta l_{p}^{2}}}\right)^{-\left(3 w_{q}+1\right)}$, and remnant mass is
 \begin{equation}
M_{r e m-E G U P}=\frac{\left(\frac{l_{p} l}{2} \sqrt{\frac{\beta}{l^{2}-\alpha \beta l_{p}^{2}}}\right)^{3}-c\left(\frac{l_{p} l}{2} \sqrt{\frac{\beta}{l^{2}-\alpha \beta l_{p}^{2}}}\right)^{-3 w_{q}+2}}{2\left(\frac{l_{p} l}{2} \sqrt{\frac{\beta}{l^{2}-\alpha \beta l_{p}^{2}}}\right)^{2}-3 g^{2}}
.\label{eq31}
\end{equation}

Next, we give a brief discussion in terms of the heat capacity function on the EGUP formalism. Firstly, the GUP modified heat capacity can be obtained when $\alpha$ is vanished
 \begin{equation}
 C_{GUP}=\frac{\pi \beta l_{p}^{2} r_{H}^{4} \sqrt{4-\frac{\beta l_{p}^{2}}{r_{H}^{2}}}\left[3 c g^{2}\left(2-3 w_{q}\right)+6 c w_{q} r_{H}^{2}-9 g^{2} r_{H}^{1+3 w_{q}}+2 r_{H}^{3+3 w_{q}}\right]}{\eta_{1}+\eta_{2}}
,\label{eq32}
\end{equation}
with
 \begin{equation}
C_{G U P}=\frac{\pi \beta l_{p}^{2} r_{H}^{4} \sqrt{4-\frac{\beta l_{p}^{2}}{r_{H}^{2}}}\left[3 c g^{2}\left(2-3 w_{q}\right)+6 c w_{q} r_{H}^{2}-9 g^{2} r_{H}^{1+3 w_{q}}+2 r_{H}^{3+3 w_{q}}\right]}{\eta_{1}+\eta_{2}}
,\label{eq33}
\end{equation}

\begin{equation}
\eta_{2}=4 r_{H}^{2}\left(-1+\sqrt{1-\frac{\beta l_{p}^{2}}{4 r_{H}^{2}}}\right)\left[27 c g^{4} w_{q}\left(2-3 w_{q}\right)+12 c g^{2} r_{H}^{2}\left(-2-3 w_{q}+9 w_{q}^{2}\right)-36 c w_{q}^{2} r_{H}^{4}+27 g^{4} r_{H}^{1+3 w_{q}}+4 r_{H}^{5+3 w_{q}}\right]
,\label{eq34}
\end{equation}

in this case we examined, a black hole remnant occurs for $r_{H}=\frac{l_{p}}{2} \sqrt{\beta}$, then the black hole remnant occurs when the remnant temperature is
\begin{equation}
T_{GUP}=\frac{1}{\pi l_{p} \sqrt{\beta}}\left[1-6 g^{2} \frac{\left(1-c 2^{3 w_{q}+1}\left(l_{p} \sqrt{\beta}\right)^{-3 w_{q}-1}\right)}{\frac{\beta l_{p}^{2}}{2}-3 g^{2}}+3 w_{q} c 2^{3 w_{q}+1}\left(l_{p} \sqrt{\beta}\right)^{-3 w_{q}-1}\right]
,\label{eq35}
\end{equation}
and remnant mass becomes
\begin{equation}
M_{rem-GUP}=\frac{\left(\frac{l_{p}}{2} \sqrt{\beta}\right)^{3}-c\left(\frac{l_{p}}{2} \sqrt{\beta}\right)^{-3 w_{q}+2}}{\frac{\beta l_{p}^{2}}{2}-3 g^{2}}
.\label{eq36}
\end{equation}

Subsequently, similarly, we express the EUP modified heat capacity function
\begin{equation}
C_{EUP}=-\frac{\left.4 l^{2} \pi r_{H}^{4} \mid 3 c g^{2}\left(2-3 w_{q}\right)+6 c w_{q} r_{H}^{2}+\left(-9 g^{2}+2 r_{H}^{2}\right) r_{H}^{1+3 w_{q}}\right]}{\lambda_{1}+\lambda_{2}}
,\label{eq37}
\end{equation}
with
\begin{equation}
\lambda_{1}=9 c g^{4} l^{2}\left(-4+9 w_{q}^{2}\right)+12 c r_{H}^{2}\left[\begin{array}{l}
g^{2} l^{2}\left(4-3 w_{q}\left(1+3 w_{q}\right)\right)+9 g^{4} w_{q} \alpha\left(-2+3 w_{q}\right) \\
+r_{H}^{2}\left(l^{2} w_{q}\left(2+3 w_{q}\right)+4 g^{2} \alpha\left(2+3 w_{q}-9 w_{q}^{2}\right)\right)+12 w_{q}^{2} \alpha r_{H}^{4}
\end{array}\right]
,\label{eq38}
\end{equation}

\begin{equation}
\lambda_{2}=r_{H}^{1+3 w_{q}}\left(27 g^{4} l^{2}-12 r_{H}^{2}\left(4 g^{2} l^{2}+9 g^{4} \alpha\right)+4 l^{2} r_{H}^{4}-16 \alpha r_{H}^{6}\right)
.\label{eq39}
\end{equation}

\begin{figure}[tbh!]
\subfigure{
\includegraphics[width=7cm]{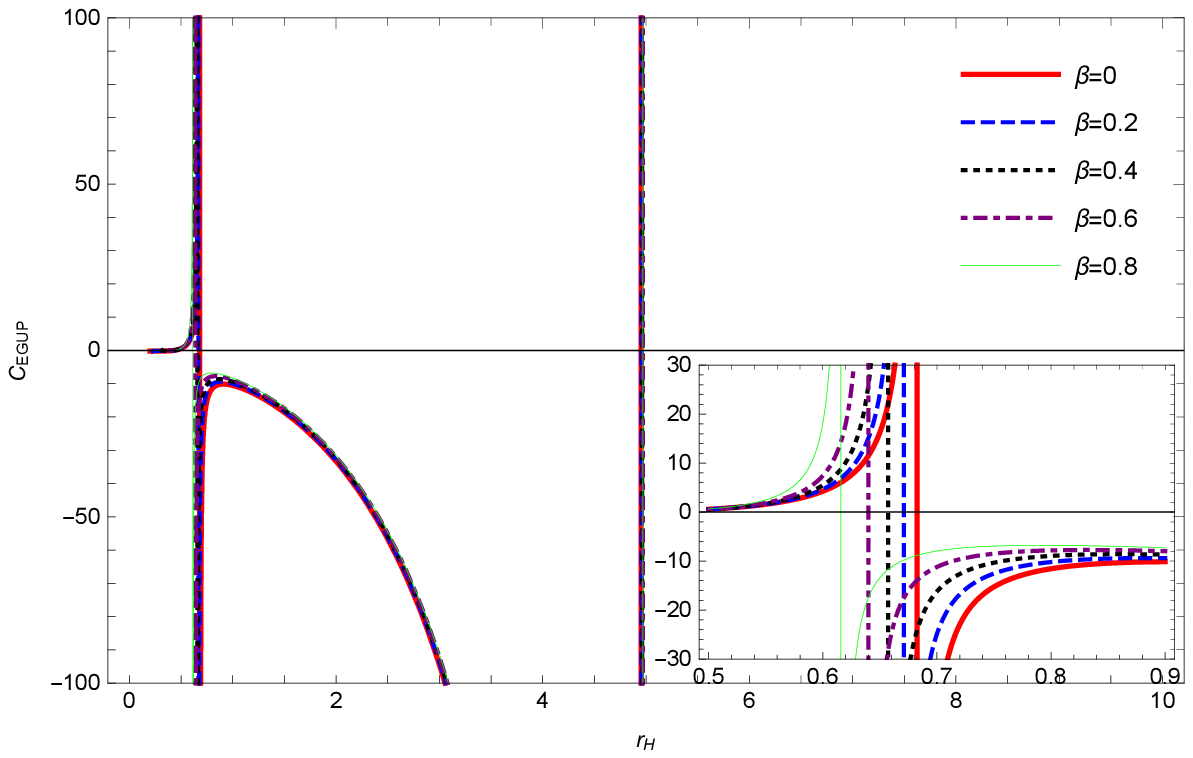}}
\subfigure{
\includegraphics[width=7cm]{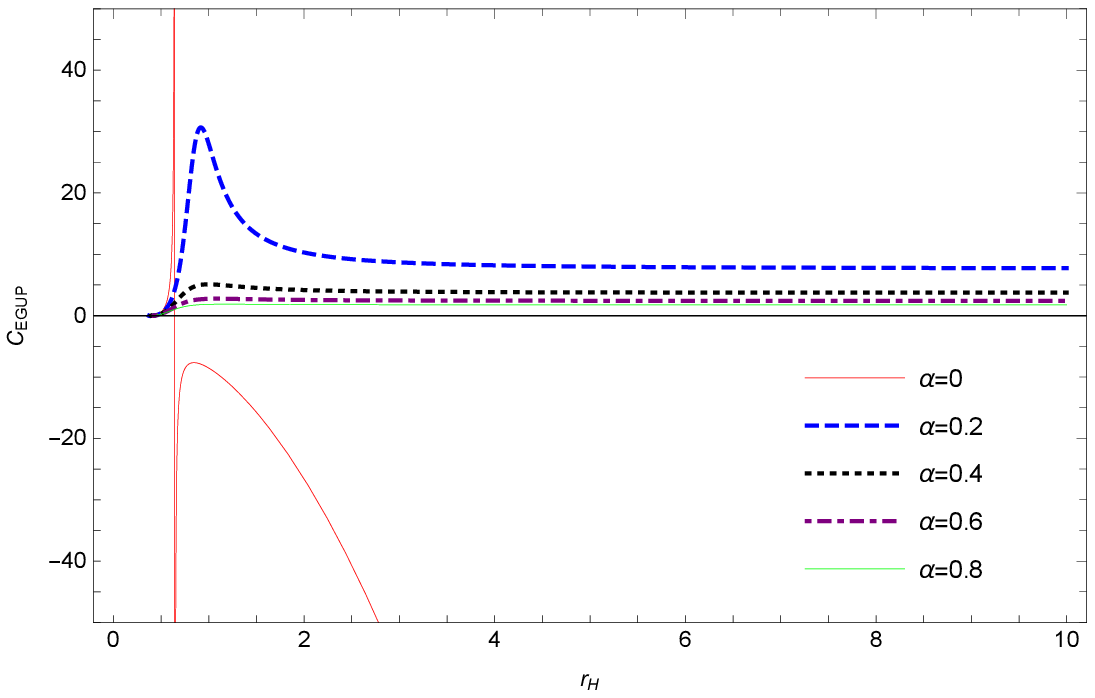}}
\caption{(left panel) The heat capacity $C_{EGUP}$ as a function of horizon radius $r_H$ for different GUP parameters, $\beta$, ($\omega_q=-1/3$, $\alpha=0.01$). (right panel) Hawking temperature $C_{EGUP}$ as a function of horizon radius $r_H$ for different EUP parameters, $\alpha$, ($\omega_q=-1/3$, $\beta=0.5$).}
\label{fig4}
\end{figure}

\begin{figure}[tbh!]
\subfigure{
\includegraphics[width=7cm]{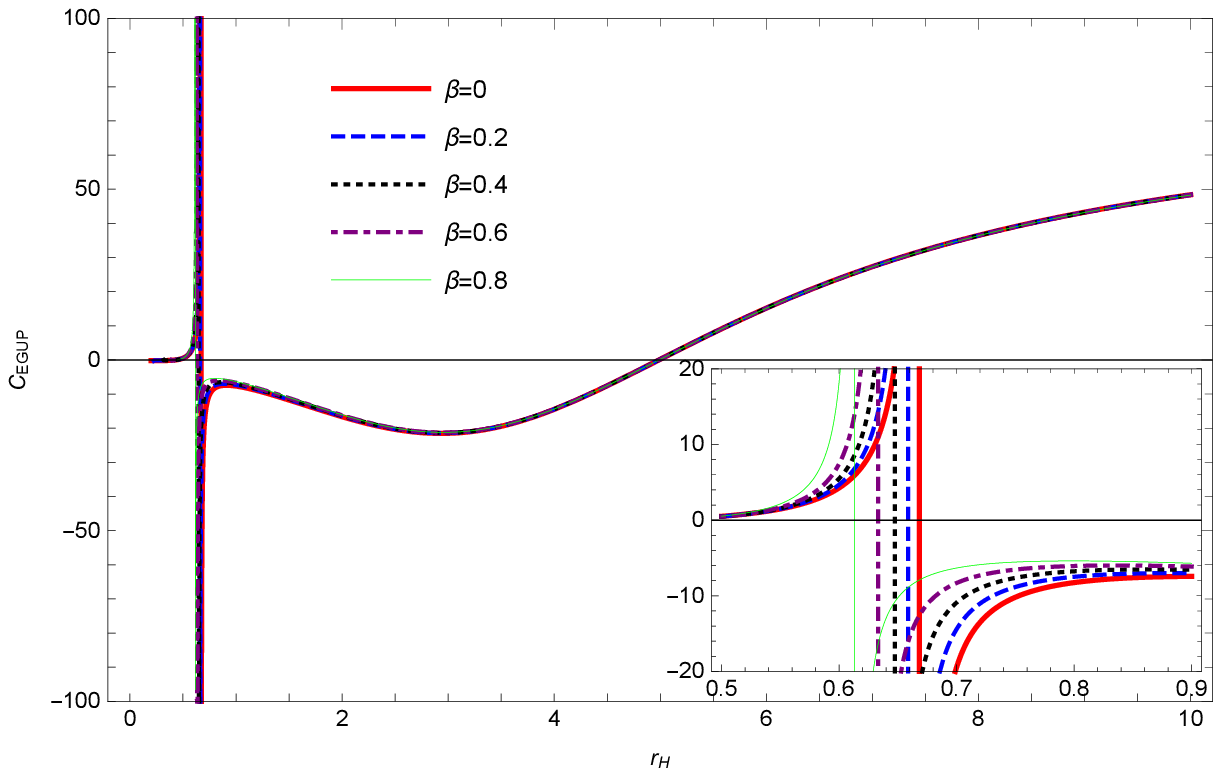}}
\subfigure{
\includegraphics[width=7cm]{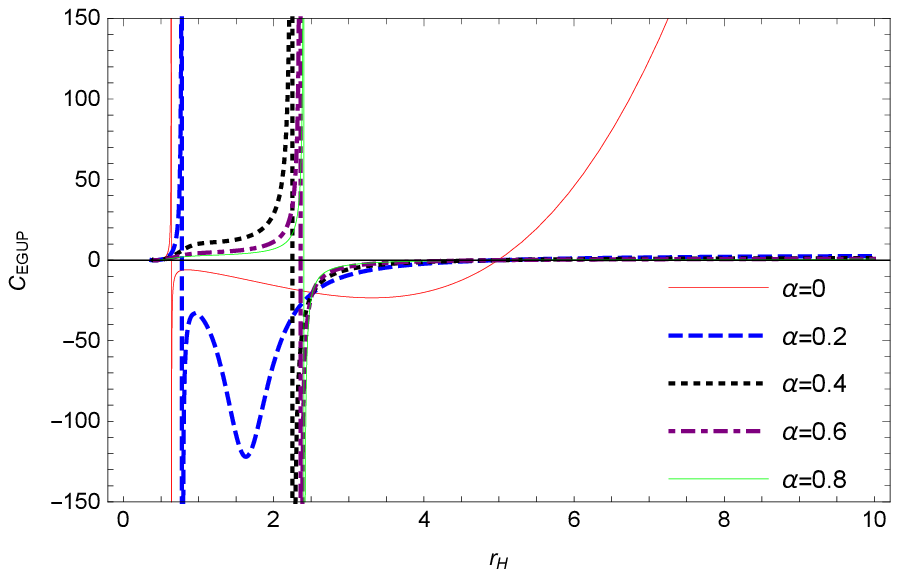}}
\caption{(left panel) The heat capacity $C_{EGUP}$ as a function of horizon radius $r_H$ for different GUP parameters, $\beta$, ($\omega_q=-2/3$, $\alpha=0.01$). (right panel) Hawking temperature $C_{EGUP}$ as a function of horizon radius $r_H$ for different EUP parameters, $\alpha$, ($\omega_q=-2/3$, $\beta=0.5$).}
\label{fig5}
\end{figure}

\begin{figure}[tbh!]
\subfigure{
\includegraphics[width=7cm]{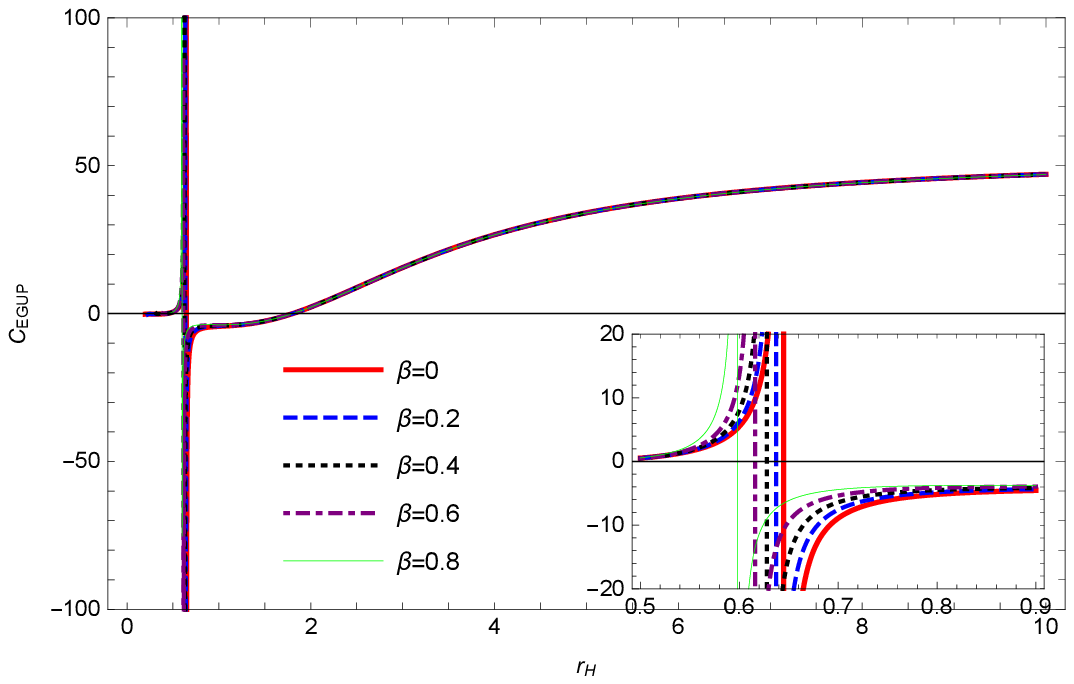}}
\subfigure{
\includegraphics[width=7cm]{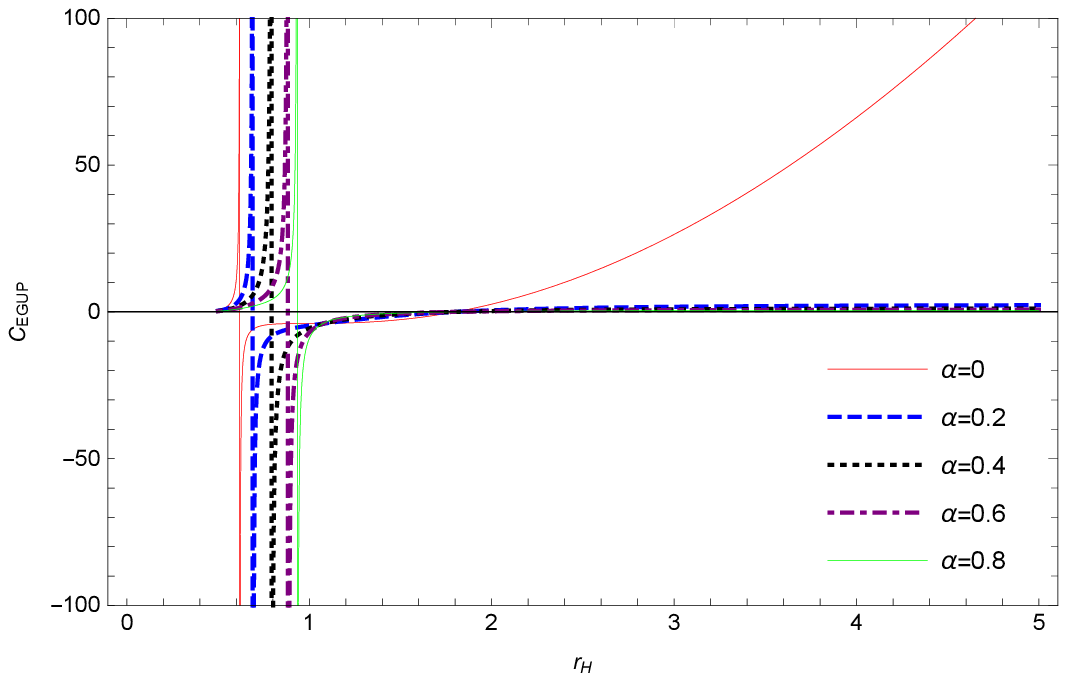}}
\caption{(left panel) The heat capacity $C_{EGUP}$ as a function of horizon radius $r_H$ for different GUP parameters, $\beta$, ($\omega_q=-1$, $\alpha=0.01$). (right panel) Hawking temperature $C_{EGUP}$ as a function of horizon radius $r_H$ for different EUP parameters, $\alpha$, ($\omega_q=-1$, $\beta=0.5$).}
\label{fig6}
\end{figure}

In order to facilitate the analysis of the effects of the EGUP parameters $\alpha$ and $\beta$ on the remnant mass, we
consider the barotropic index as $\omega_{q}=-1/3$, $\omega_{q}=-2/3$ and $\omega_{q}=-1$ in Table \protect \ref{table5}. For all three cases, which show that when the GUP parameter remain a constant, the remnant mass increases with the increasing EUP parameter. Similarly, if we takes a constant EUP parameter, the remnant mass increases with the increasing GUP parameter. And in order to illustrate the EGUP modified heat capacity versus horizon visually, we depict them in Fig.\protect \ref{fig4}, Fig.\protect \ref{fig5} and Fig.\protect \ref{fig6} for $\omega_q=-1/3$, $\omega_q=-2/3$ and $\omega_q=-1$ respectively. In Fig.\protect \ref{fig4} (left panel), by remaining a constant EUP parameter $\alpha=0.01$, the barotropic index $\omega_{q}=-1/3$, and three GUP parameters, $\beta=0.0, 0.2, 0.4, 0.6, 0.8$, we find that the trend effect of the different GUP on the heat capacity is very weak. While for the different EUP parameters, our result indicates that the completely different trend of the heat capacity, which implies that the EUP parameter $\alpha$ has a certain effect in Fig.\protect \ref{fig4}. Similarly, the case of $\omega_{q}=-2/3$ and $\omega_{q}=-1$ can also be found in Fig.\protect \ref{fig5} and Fig.\protect \ref{fig6}.

\begin{table*}
\begin{tabular}{|c|c|c|c|c||c|}
\hline
$M_{rem}$ \\
        \hline
\hline
$l=1$& $l_{p}=1$& $g=0.1$& $c=0.1$\\
\hline
$\omega_{q}=-1/3$ \\

\hline
\hline
$\alpha$ &  $\beta=0.2$& $\beta=0.4$ & $\beta=0.6$ &  $\beta=0.8$  & $\beta=1$   \\
\hline
\hline
$0.0$ & 0.143747  & 0.167415& 0.193649 & 0.217563 & 0.239362 \\
$0.2$& 0.144239  &0.172112  &  0.203715 & 0.234341 & 0.264241  \\
$0.4$ & 0.144899  & 0.177648& 0.216361 & 0.257162& 0.301321 \\
$0.6$ & 0.14574 & 0.184235  &0.232751 &0.290404  & 0.364504 \\
$0.8$&  0.146776  & 0.192168   &0.254946  &0.344718   & 0.509226  \\

\hline
$\omega_{q}=-2/3$ \\
\hline
\hline
$0.0$ & 0.156148& 0.180134& 0.206832& 0.230926& 0.25266\\
$0.2$& 0.156606 & 0.184931& 0.217004& 0.247674& 0.277189\\
$0.4$& 0.157245 & 0.190577& 0.229721& 0.270239& 0.31319\\
$0.6$& 0.158073  & 0.19728& 0.246093& 0.30266& 0.372986\\
$0.8$& 0.159106 & 0.205332& 0.268059& 0.354471& 0.502548\\
\hline
$\omega_{q}=-1$ \\
\hline
\hline
$0.0$ & 0.158921& 0.184156& 0.211938& 0.236902& 0.259309\\
$0.2$& 0.159431 & 0.189157& 0.222491& 0.254179& 0.284426\\
$0.4$& 0.160124 & 0.195037& 0.235657& 0.277332& 0.320851\\
$0.6$& 0.161013  & 0.202012& 0.252551& 0.310261& 0.379692\\
$0.8$& 0.162114 & 0.21038& 0.275102& 0.361741& 0.495081\\

\end{tabular}
\caption{the remnant mass as a function for the different EGUP parameters $\beta$ and $\alpha$.}\label{table5}
\end{table*}

Next, we express the EGUP modified entropy function of the black hole as
\begin{equation}
S_{EGUP}=-\frac{l \pi}{8 \alpha\left(l^{2}+6 g^{2} \alpha\right)^{[}}\left[-2 l^{3} \ln \left(2-\chi_{0}\right)+\chi_{1}+\chi_{2}+\chi_{3}\right]
,\label{eq40}
\end{equation}
with
\begin{equation}
\chi_{0}=\sqrt{4+l_{p}^{2}\left(-\frac{\beta}{r_{H}^{2}}-\frac{4 \alpha \beta}{l^{2}}\right)}
,\label{eq41}
\end{equation}

\begin{equation}
\chi_{1}=\left(l^{2}+6 g^{2} \alpha\right)\left[\left(l+\sqrt{l^{2}-l_{p}^{2} \alpha \beta}\right) \ln \left(2 \sqrt{l^{2}-l_{p}^{2} \alpha \beta}-l \chi_{0}\right)+\left(l-\sqrt{l^{2}-l_{p}^{2} \alpha \beta}\right) \ln \left(2 \sqrt{l^{2}-l_{p}^{2} \alpha \beta}+l \chi_{0}\right)\right]
,\label{eq42}
\end{equation}

\begin{equation}
\chi_{2}=-g \alpha\left[6 g l+\sqrt{6} \sqrt{-6 g^{2} l_{p}^{2} \alpha \beta+l^{2}\left(6 g^{2}-l_{p}^{2} \beta\right)} \ln \left(\sqrt{2} \sqrt{-6 l_{p}^{2} g^{2} \alpha \beta+l^{2}\left(6 g^{2}-l_{p}^{2} \beta\right)}-\sqrt{3} g l \chi_{0}\right)\right]
,\label{eq43}
\end{equation}

\begin{equation}
\chi_{3}=-g \alpha\left[6 g l-\sqrt{6} \sqrt{-6 g^{2} l_{p}^{2} \alpha \beta+l^{2}\left(6 g^{2}-l_{p}^{2} \beta\right)} \ln \left(\sqrt{2} \sqrt{-6 l_{p}^{2} g^{2} \alpha \beta+l^{2}\left(6 g^{2}-l_{p}^{2} \beta\right)}+\sqrt{3} g l \chi_{0}\right)\right]
.\label{eq44}
\end{equation}
Subsequently, we present a brief discussion in terms of the modified entropy function, i.e a taylor expansion for this expression

\begin{equation}
S_{EGUP}=\pi r_{H}^{2}-\frac{2 r_{H}^{2} \alpha \pi}{l^{2}}\left(3 g^{2}+r_{H}^{2}\right)+\frac{\pi}{4 l^{2} \alpha}\left(6 \alpha g^{2}-l^{2}\right)^{2} \ln \left(\frac{l_{p}^{2} \beta}{4 r_{H}^{2}}\right)+O[\alpha]^{2}+O[\beta]^{2}
,\label{eq45}
\end{equation}
it can be observed that quintessence matter will not affect the entropy of the Bardeen Black Hole.

Finally, we present the pressure $P_{q}$ and matter-energy density $\rho_{q}$ of the Bardeen Black Hole with quintessence matter as
\begin{equation}
P_{q}=w_{q} \rho_{q}=-\frac{3 c}{2} \frac{w_{q}^{2}}{r^{3\left(1+w_{q}\right)}}
.\label{eq46}
\end{equation}
and in this case, considering the volume $P_{q}=w_{q} \rho_{q}=-\frac{3 c}{2} \frac{w_{q}^{2}}{r^{3\left(1+w_{q}\right)}}$ in terms of the radius \cite{ret32}, we rewrite the Hawking temperature as

\begin{equation}
\left.\mathrm{T}_{EGUP}=\frac{2\left(3 V w_{q}^{2}\right)^{\frac{1}{3}}}{\pi \beta l_{p}^{2}}\left[1-6 g^{2} \frac{\left (1+\frac{2 P_{q}\left(3 V w_{q}^{2}\right)^{\frac{2}{3}}}{3 w_{q}^{2}}\right.)}{2\left(3V w_{q}^{2}\right)^{\frac{2}{3}}-3 g^{2}}-\frac{2 P_{q}\left(3 V w_{q}^{2}\right)^{\frac{2}{3}}}{w_{q}}\right][ 1-\sqrt{1-\beta l_{p}^{2}\left(\frac{1}{4\left(3 V w_{q}^{2} \frac{2^{2}}{3}\right.}+\frac{\alpha}{l^{2}}\right)}\right],
\label{eq47}
\end{equation}
then if we select T=1 isotherm, the EGUP-corrected equation of state of the Bardeen black hole becomes
\begin{equation}
P_{q-E G U P}=\frac{2 V w_{q}^{2}-3^{\frac{4}{3}} g^{2}\left(V w_{q}^{2}\right)^{\frac{1}{3}}+\left[\pi \beta l_{p}^{2}\left(-3 g^{2}+2 \times 3^{\frac{2}{3}}\left(V w_{q}^{2}\right)^{\frac{2}{3}}\right) /-6+\sqrt{36-\beta l_{p}^{2}\left(\frac{3^{\frac{4}{3}}}{\left(V w_{q}^{2}\right)^{\frac{2}{3}}}+\frac{36 \alpha}{l^{2}}\right)}\right]}{2 V\left[g^{2}\left(2-3 w_{q}\right)+2 \times 3^{\frac{2}{3}} w_{q}\left(V w_{q}^{2}\right)^{\frac{2}{3}}\right]}.
\label{eq48}
\end{equation}

Now, we give a brief discussion of the EGUP-corrected Pressure-Volume (P-V) isotherm for the black hole with quintessence matter. For example, the GUP-corrected pressure of black hole for $\alpha=0$ is
\begin{equation}
P_{q-G U P}=\frac{1}{2 V\left[g^{2}\left(2-3 w_{q}\right)+2 \times 3^{\frac{2}{3}} w_{q}\left(V w_{q}^{2}\right)^{\frac{2}{3}}\right]}\left[V w_{q}^{2}-3^{\frac{4}{3}} g^{2}\left(V w_{q}^{2}\right)^{\frac{1}{3}}+\frac{\pi \beta l_{p}^{2}\left(-3 g^{2}+2 \times 3^{\frac{2}{3}}\left(V w_{q}^{2}\right)^{\frac{2}{3}}\right)}{\left.-6+\sqrt{36-\frac{3^{\frac{4}{3}} \beta l_{p}^{2}}{\left(V w_{q}^{2}\right)^{\frac{2}{3}}}}\right]}\right]. \label{eq49}
\end{equation}
and For $\beta=0$, the EUP-corrected pressure of black hole reads
\begin{equation}
P_{q-E U P}=\frac{l^{2}\left[2 V w_{q}^{2}\left(1-4 \times 3^{\frac{1}{3}} \pi\left(V w_{q}^{2}\right)^{\frac{1}{3}}\right)+3^{\frac{1}{3}} g^{2}\left(V w_{q}^{2}\right)^{\frac{1}{3}}\left(-3+4 \times 3^{\frac{1}{3}} \pi\left(V w_{q}^{2}\right)^{\frac{1}{3}}\right)\right]+\zeta}{2V\left[g^{2}\left(2-3 w_{q}\right)+2 \times 3^{\frac{2}{3}} w_{q}\left(V w_{q}^{2}\right)^{\frac{2}{3}}\right]\left[l^{2}+4 \times 3^{\frac{2}{3}} \alpha\left(V w_{q}^{2}\right)^{\frac{2}{3}}\right]}
\label{eq50}
\end{equation}
with $\zeta=4 V w_{q}^{2} \alpha\left(-9 g^{2}+2 \times 3^{\frac{2}{3}}\left(V w_{q}^{2}\right)^{\frac{2}{3}}\right)$,
while for $\beta=0$ and $\alpha=0$, Eq.\protect \eqref{eq48} reduces to HUP-modified pressure as
\begin{equation}
P_{q-HUP}=\frac{w_{q}^{2}\left[8 \times 3^{\frac{2}{3}} \pi V w_{q}^{2}-2 \times 3^{\frac{1}{3}}\left(V w_{q}^{2}\right)^{\frac{2}{3}}+3 g^{2}\left(3^{\frac{2}{3}}-4 \pi\left(V w_{q}^{2}\right)^{\frac{1}{3}}\right)\right]}{2 \times 3^{\frac{1}{3}} g^{2}\left(2-3 w_{q}\right)\left(V w_{q}^{2}\right)^{\frac{2}{3}}+12 w_{q}\left(V w_{q}^{2}\right)^{\frac{4}{3}}}.
\label{eq51}
\end{equation}

\begin{figure}[tbh!]
\subfigure{
\includegraphics[width=7cm]{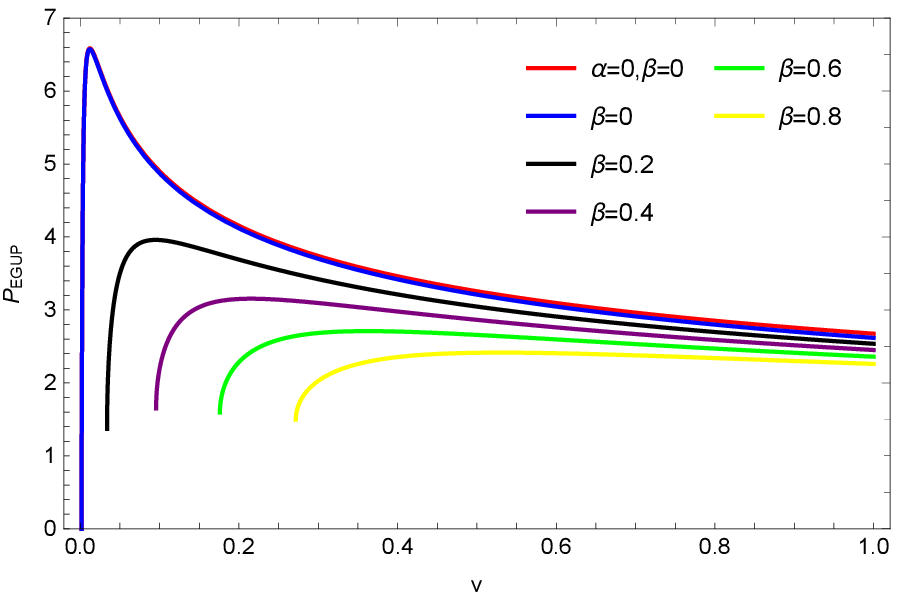}}
\subfigure{
\includegraphics[width=7cm]{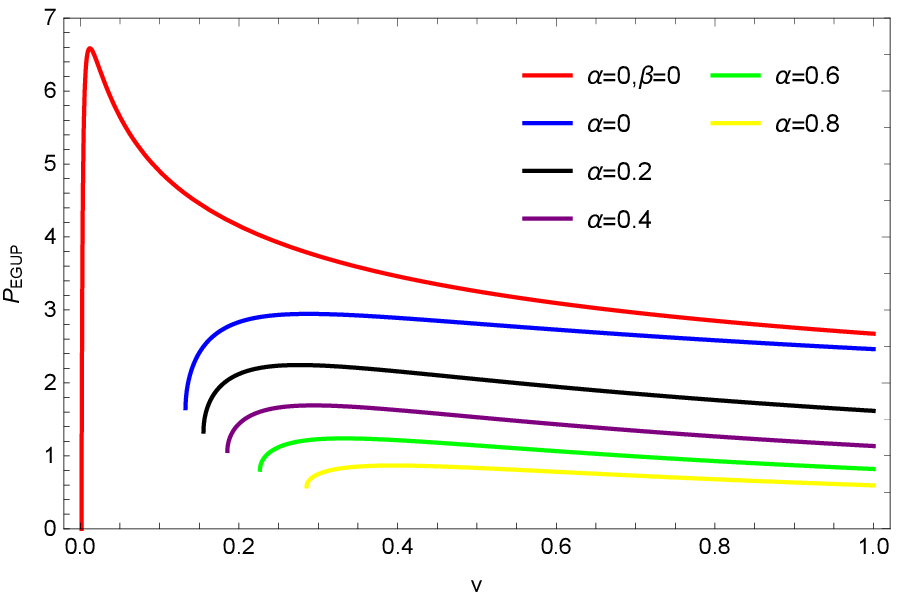}}
\caption{(left panel) The EGUP-corrected pressure as a function of the volume $V$ for different GUP parameters, $\beta$, ($\omega_q=-1$, $\alpha=0.01$). (right panel) The EGUP-corrected pressure as a function of the volume $V$ for different EUP parameters, $\alpha$, ($\omega_q=-1$, $\beta=0.5$).}
\label{fig7}
\end{figure}

\begin{figure}[tbh!]
\subfigure{
\includegraphics[width=7cm]{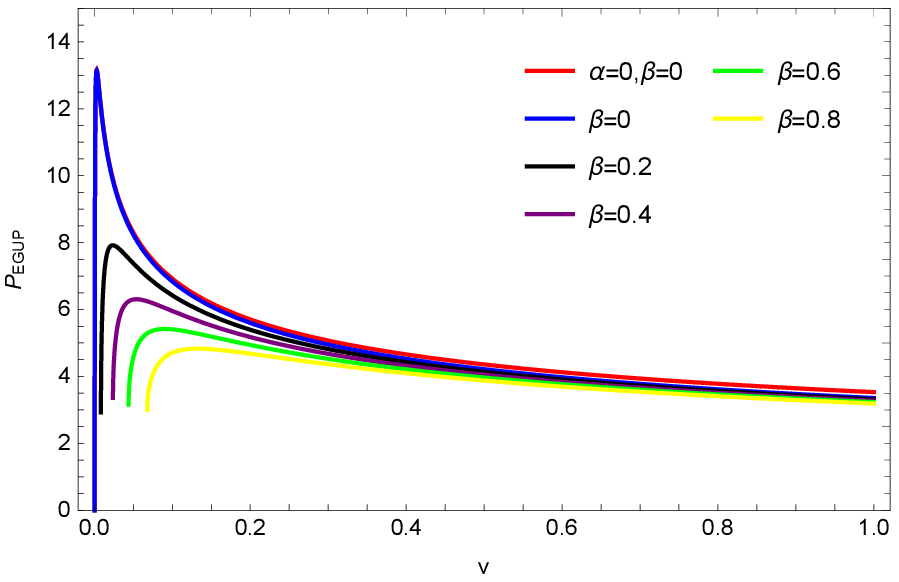}}
\subfigure{
\includegraphics[width=7cm]{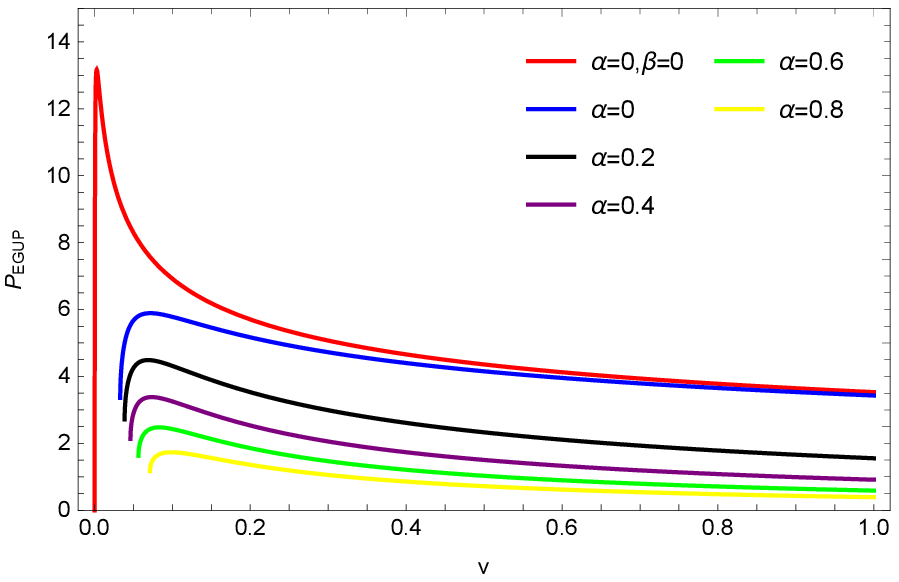}}
\caption{(left panel) The EGUP-corrected pressure as a function of the volume $V$ for different GUP parameters, $\beta$, ($\omega_q=-1$, $\alpha=0.01$). (right panel) The EGUP-corrected pressure as a function of the volume $V$ for different EUP parameters, $\alpha$, ($\omega_q=-1$, $\beta=0.5$).}
\label{fig8}
\end{figure}

\begin{figure}[tbh!]
\subfigure{
\includegraphics[width=7cm]{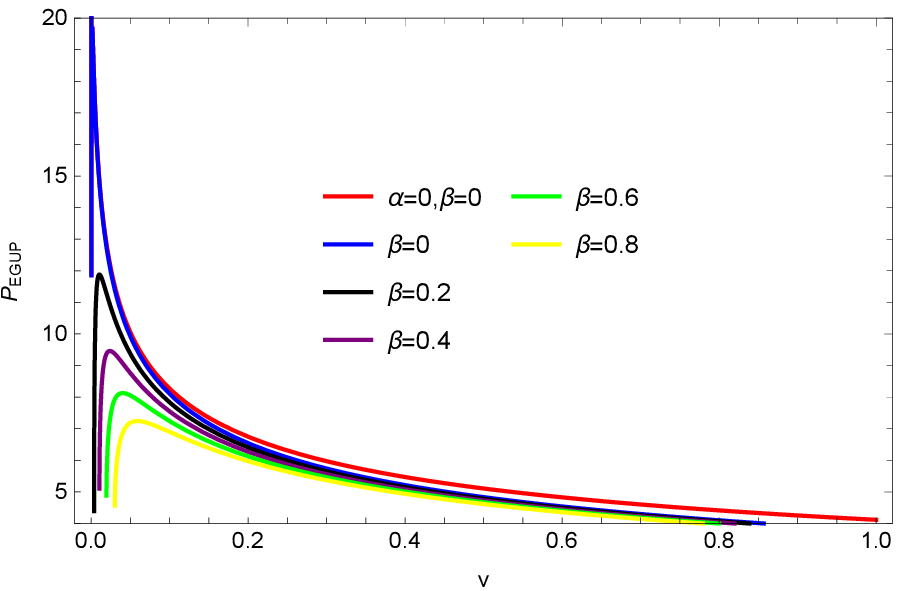}}
\subfigure{
\includegraphics[width=7cm]{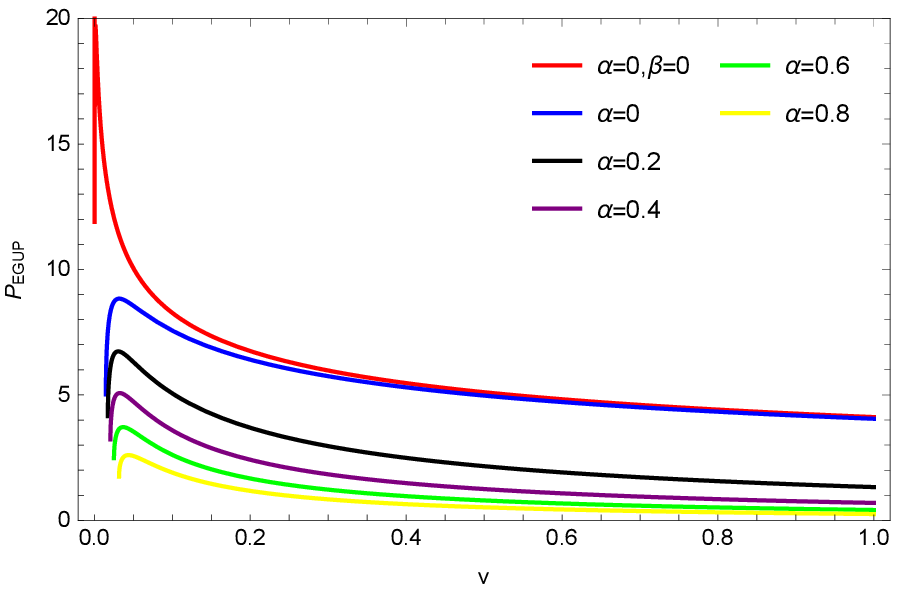}}
\caption{(left panel) The EGUP-corrected pressure as a function of the volume $V$ for different GUP parameters, $\beta$, ($\omega_q=-1$, $\alpha=0.01$). (right panel) The EGUP-corrected pressure as a function of the volume $V$ for different EUP parameters, $\alpha$, ($\omega_q=-1$, $\beta=0.5$).}
\label{fig9}
\end{figure}

Here, in order to analyze the effect of EGUP parameters $\beta, \alpha$ on the P-V isotherm, we illustrate graphically  the modified P-V isotherm for the barotropic index $\omega_{q}=-1/3$, $\omega_{q}=-2/3$ and $\omega_{q}=-1$ in Fig.\ref{fig7}, Fig.\ref{fig8} and Fig.\ref{fig9} respectively. For all three cases, we can see that if the volume $V$ remains a constant, the EGUP-corrected pressure will decrease with the increasing GUP parameter or EUP parameter.

\section{Conclusions} \label{sec4}
In this letter, we examined the thermodynamics of the Bardeen Black Hole with quintessence matter by employing the extended generalized uncertainty principle. Firstly, we briefly review the new extended uncertainty principle which is a summation of generalized and extended uncertainty principles. Subsequently, the quantum corrected-Hawking temperature of the Bardeen Black Hole with quintessence matter is obtained. In order to facilitate the analysis of the nature of the accelerated expansion of the universe and quintessence regime of dark energy, we illustrate graphically the barotropic index $\omega_{q}$ as $-1/3$, $-2/3$ and $-1$, respectively, which show that different values of the barotropic index will lead to different areas of the effective horizon radius, and the EGUP parameters result in a lower bound on the horizon. Moreover, we depict the effect of the GUP and EUP parameters in terms of the characteristic of the horizon range and Hawking temperature in Fig.\protect \ref{fig1}-Fig.\protect \ref{fig3} and Table \protect \ref{table1}-Table \protect \ref{table4}. Subsequently, we discuss the modified remnant mass, heat capacity, and entropy functions, which show that when the GUP parameter remain a constant, the remnant mass increases with the increasing EUP parameter, and if we takes a constant EUP parameter, the remnant mass increases with the increasing GUP parameter. While the quintessence matter will not affect the entropy of the Bardeen Black Hole. Finally, we present a detailed analysis of the EGUP modified P-V isotherm graphically.

\section*{Acknowledgments}
This research was funded by the Guizhou Scientific Foundation-ZK[2022] General 491 and the National Natural Science Foundation of China (Grant No.11465006).



\end{document}